\documentclass[pre,reprint]{revtex4-1}
\usepackage{amsmath}
\usepackage{amsfonts}
\usepackage{amssymb}
\usepackage{verbatim}
\usepackage{float}
\usepackage{tikz}
\def\mean#1{\left \langle#1\right \rangle}

\usepackage{natbib}

\begin{document}

\title[JASA]{Vocal wow in an adapted reflex resonance model}

\author{Fran\c{c}ois-Xavier Brajot }
\email{brajot@ohio.edu}
\affiliation{Communication Sciences and Disorders, Ohio University, Grover Center W221, Athens, Ohio 45701, USA}
	
\author{Alexander B. Neiman}
\email{neimana@ohio.edu}
\affiliation{Department of Physics and Astronomy, Ohio University, Athens, Ohio 45701, USA}
\affiliation{Neuroscience Program, Ohio University, Athens, Ohio 45701, USA}

\date{\today}

\begin{abstract} 
Vocal wow is a slow 0.2 - 3 Hz modulation of the voice that may be distinguished from the 4 - 7 Hz modulation of tremor or vibrato. We use a simple model of laryngeal muscle activation, mediated by time-delayed auditory feedback, to show that wow may arise due to Andronov-Hopf bifurcation.The model demonstrates a differential effect of feedback gain and delay on modulation depth and frequency, respectively. Parametric formulas for recovering feedback parameters from the acoustic signal are presented. Interactions between reflex and auditory parameters are also assessed in a full model that includes a neuromuscular reflex loop. Model predictions are tested in two subjects.
\end{abstract}

\pacs{43.25.Ts, 43.66.Hg, 43.70.Bk} 


\maketitle

\section{Introduction}
Long-term phonatory instabilities are slow fluctuations of the voice distinguished by how rapidly they modulate pitch or intensity. The singer's vibrato, for example, is a consciously controlled 4-7 Hz fluctuation of vocal fundamental frequency ($f_o$), similar in many respects to the common vocal tremor. Borrowing terms from the audio recording industry, \citet{ternstrom1989} differentiated slow and fast vibrato as "wow" and "flutter," respectively. 
The rationale for this distinction is that each type of fluctuation may result from different neurophysiological mechanisms.
In particular, the more rapid fluctuations are associated with proprioceptive feedback, whereas the slower vocal wow is attributed to the auditory feedback loop. 

Clinical observation of vocal tremors has prompted a refinement in terminology to a tripartite distinction, with wow below 2 or 3 Hz, tremor between 4 and 7 Hz, and flutter extending roughly from 8 to 20 Hz \citep{aronson1992, hartelius1997}. 
This distinction has proven relevant for the differential diagnosis of neurogenic voice disorders and could be useful in further distinguishing subtypes \citep{buder2003}. 
Disorders with ostensible somatosensory feedback deficits such as Parkinson disease are typically associated with vocal tremor, for example \citep{conte2015, hammer2010}.
Vocal wow, on the other hand, is more readily observed in disorders that affect auditory feedback, such as sensorineural hearing loss \citep{lee2009}. 

Attributing long-term phonatory instabilities to sensorimotor feedback loops places their clinical manifestation within the purview of dynamical diseases \citep{mackey1977, mackey1987dynamical}.
A wide range of physiological behaviors are periodic in nature and many disorders are characterized by a shift to a new periodic regime.
In sensorimotor control, these represent a transition in system dynamics as the characteristic time or gain in a feedback loop are pushed beyond some threshold \citep{glass1988delays}.
The challenge lies in identifying those parameters most relevant for promoting the transition.

The reflex resonance model proposed by \citet{titze2002reflex} has been an important step in that direction.
Developed to explain vocal vibrato, it has a clear biophysiocal basis and draws on well-established properties of negative feedback systems to oscillate when adjustments are made to feedback gain or conduction times.
The model explains a number of empirical observations, including greater control over vibrato extent than frequency, age-related variability and possible origins to certain pathological tremors. 
Importantly, it provides an excellent basis from which to assess the putative role of other feedback loops in the generation of phonatory instabilities, namely that of auditory feedback and vocal wow.

In a recent experimental study, vocal wow was elicited by delaying speakers' auditory feedback during sustained phonation \citep{brajot2018delay}. The predominant oscillation in $f_o$ was consistently below 2 Hz and decreased as feedback delay increased. The depth of modulation also appeared to increased with larger delays, but this varied considerably across subjects.
Possible effects of delayed auditory feedback on higher frequency modulations were not evaluated.
In order to elicit a natural response, moreover, participants were not instructed to maintain a steady intensity and feedback gain was not controlled for.
The frequency resolution of analyzed $f_o$ contours was also inherently limited by participants' maximum phonation times, which often did not extend beyond 10 seconds.
These limitations motivated the present effort to formally model the phenomenon.  

The principal objectives of the current analysis were therefore (1) to determine whether adding an auditory feedback loop to the reflex resonance model could approximate existing empirical findings on delay-induced wow and, if so, (2) to determine whether system parameters could be derived from the output and (3) characterize possible interactions between wow and vibrato.

\section{Model}
We begin by summarizing the system proposed by Titze et al. \citep{titze1995coupling,titze2002reflex}.
Both, the cricothyroid (CT) and thyroarytenoid (TA) muscles are described by a Kelvin model, with some simpifcations resulting in a 2D system for the internal contractile stress, $\sigma$, and the force generated by the muscle, $F$. The model's equations are
\begin{eqnarray}
\label{model1.eq}
&& t_\sigma \dot{\sigma} = \sigma_A(t) - \sigma, \nonumber \\
&& t_s \dot{F} = S \sigma - F, 
\end{eqnarray}
where $t_\sigma$ is the activation time constant, $t_s$ is the contraction time constant,  $S$ is the cross-sectional area of the muscle. The same equations, but with different constants, are used for the CT and TA muscles.
In Eqs.(\ref{model1.eq}), $\sigma_A (t)$ is the fully developed active stress, which can be written as
$\sigma_A(t)=a(t) \sigma_\text{max}$, where $\sigma_\text{max}$ is the maximum active stress and $a(t)$ is a dimensionless activation variable. 

In the original model, $a(t)$ is modulated by the time-delayed variations of the vocal-fold lengths, quantified in terms of the vocal-fold strains of TA and CT agonist-antagonist muscle pair. A strong enough magnitude of the reflex feedback results in 5--7 Hz oscillation of the vocal fundamental requency, $fo$.
Here, in addition to the reflex feedback, we use auditory time-delayed feedback. We first describe the dynamics of the model with auditory feedback only and demonstrate the emergence of low-frequency wow oscillations. For this, equations for TA  muscle only suffice. Later, we incorporate both the auditory and reflex feedbacks and discuss quasiperiodic oscillations.

In the oscillatory form, Eqs.(\ref{model1.eq}) become,
\begin{equation}
\label{model2.eq}
\ddot{F}_k + \mu_k \dot{F}_k +\Omega_k^2 F = \Omega_k^2 S_k \sigma_{\text{max},k} \, a_k(t), 
\end{equation}
where the subscript $k$ is used to distinguish CT ($k=1$) and TA ($k=2$) muscles;
$\Omega_k^2=1/(t_{\sigma, k}\, t_{s k})$ and $\mu_k=(t_{\sigma, k}+t_{s, k})\Omega_k^2$.  

Eqs.(\ref{model2.eq})  for forces are augmented by two 2-nd order nonlinear differential equations for translation and rotation of the cricoid cartilage. We have not modified these equations and refer to Eqs.(8--11) and corresponding parameters in \citet{titze2002reflex}. 
The activations variables $a_k(t)$ in (\ref{model2.eq}) are modulated by the time delayed strains of CT and TA muscles and their rates, $\varepsilon_k(t)$, $\dot{\varepsilon}_k(t)$,  calculated from the translation and rotation of  the cricoid cartalage [Eqs.(10,11) in \citet{titze2002reflex}].  In addition, a broad-band noise is added to the cortically generated activation to laryngeal muscles \citep{titze2002reflex}. This ambient muscle excitation from the CNS was modeled by Gaussian noise, $\xi(t)$, band-limited to 0 -- 15 Hz with the standard deviation (SD) $q$, and the power spectral density (PSD), 
$P_{\xi}(f)=q^2/(2f_c)$, for $ |f| \le f_c=15$~Hz and 0, otherwise.
The activation variables become,
\begin{equation}
a_k(t)=a_{o,k}\left\{ 1+g_r [\varepsilon_k(t-\tau_r)+ u_r \dot{\varepsilon}_k(t-\tau_r) ] + q \xi(t) \right\},
\label{active1.eq}
\end{equation}
where $a_{o,k}$ are constant values of the activation, $g_r$ is the reflex feedback strength, $\tau_r$ is the reflex feedback delay, and $u_r$ is the scaling factor for the strain rate. 

The time-dependent vocal fundamental frequency, i.e. the $f_o$ contour, is  obtained using the vibrating string formula,
\begin{equation}
\label{Fo.eq}
f_o(t) = \frac{1}{2L_0}\sqrt{\frac{F_2(t)}{\rho S_2}},
\end{equation}
where $F_2(t)$ is the force generated by the TA muscle, $\rho=1140$ kg/m$^3$ is the tissue density, $L_0=18.3$~mm is the rest TA muscle length.

In the following we use the same values for biomechanical constants as in
Tables I and II in \citet{titze2002reflex} for the TA muscle:
$t_\sigma=0.01$~s, $t_s=0.044$~s, $\sigma_\text{max}=105$~kPa, $S=40.9$~mm$^2$; and
for CT muscle: 
$t_\sigma=0.01$~s, $t_s=0.09$~s, $\sigma_\text{max}=89$~kPa, $S=73.8$~mm$^2$.

\subsection{Linear model with auditory delayed feedback}

In this section we consider the linear model without the reflex feedback, $g_r=0$ in (\ref{active1.eq}). We consider the TA muscle only, $k=2$, and thus drop subscript index $k$ in (\ref{model2.eq}) and (\ref{active1.eq}).
For a constant value of activation, $a(t)=a_0$, the equilibrium is
\begin{equation}
\label{Feq.eq}
F_0= S \sigma_\text{max} \,  a_0.
\end{equation}
For the biomechanics parameter values $\mu$ and $\Omega$, the oscillator (\ref{model2.eq}) is in the overdamped regime.
We set the equilibrium value of the force generated by the TA muscle as $F_0=1$~N, which gives $a_0=0.233$.
\begin{figure*}[ht]
	\centering 
	\includegraphics[width=\textwidth]{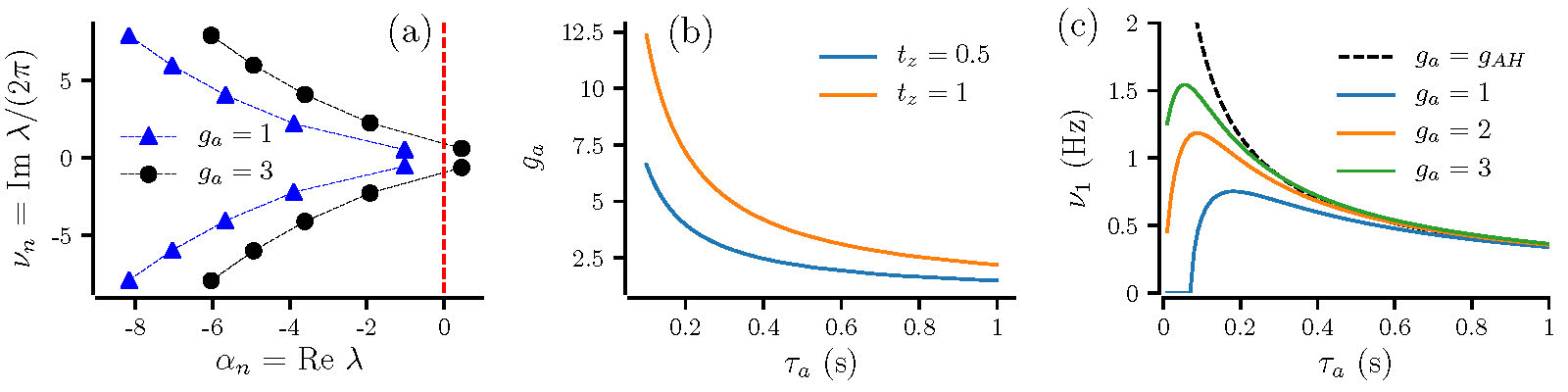}
	\caption{(color online) Stability of equilibrium of the linear model. (a): Complex conjugate eigenvalues for the indicated values of the feedback strength, $g_a$. Other parameters are: $t_z=0.5$, $\tau_a=0.5$ s.
		(b): Lines of Andronov-Hopf bifurcation on the parameter plane $(\tau_a, g_a)$ for the indicated values of $t_z$ obtained from Eqs.(\ref{tau_hopf.eq},\ref{g_hopf.eq}).
		(c): Lowest frequency, $\nu_1$, versus the delay time for the indicated values of the feedback strength. Dashed line shows the dependence of $\nu_1 (\tau_a)$ at the bifurcation line, given by Eq.(\ref{tau_hopf.eq}) with $\beta=2\pi\nu_1$ and $t_z=0.5$~s.  
	}
	\label{eigen.fig}
\end{figure*}

\begin{figure}
	\centering 
	\includegraphics[width=0.45\textwidth]{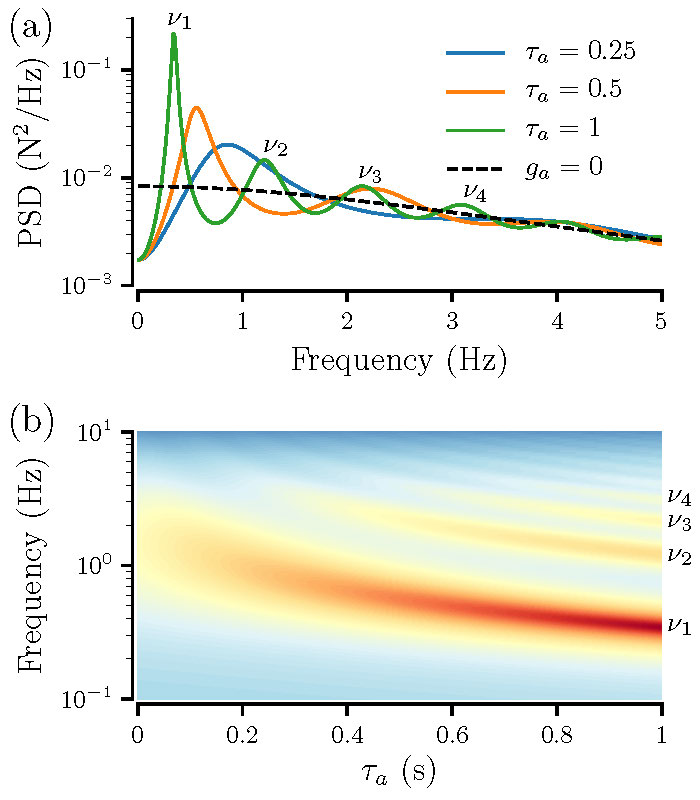}
	\caption{(color online) Power spectral density (PSD) (\ref{psd_th.eq}) of the linear model. (a): PSD for the indicated values of delay time (in sec). Dashed line shows the PSD with no feedback, $g_a=0$, according to Eq.(\ref{PSD_nofbk.eq}).
		(b): Heat map of the PSD vs time delay, $\tau_a$. 
		Peaks in the PSD occurs at frequencies corresponding to the imaginary parts of eigenvalues, $\nu_n$, marked for the  $\tau_a=1$~s curve on (a) and to the right of the heat map on panel (b).
		Other parameters are: $t_z=0.5$~s, $g_a=1.2$, $q=0.5$.} 
	\label{psd_th.fig}
\end{figure}

The power spectral density (PSD) of the force fluctuations around the equilibrium, $F_0$, $P_{F}(f)$ can be obtained  from (\ref{model2.eq}) by calculating the Fourier transform of the force, $\tilde{F}(f)$ and ensemble averaging over noise realizations,
$P_F (f)=\mean{|\tilde{F}(f)|^2}$,
yielding,
\begin{equation}
\label{PSD_nofbk.eq}
P_F(f)=\frac{q^2\Omega^4 S^2\sigma^2_\text{max}}{2f_c\left[(4\pi^2 f^2-\Omega^2)^2+4\pi^2 \mu^2 f^2\right]}.
\end{equation}
The PSD peaks at zero frequency and shows no other peaks [dashed line in Fig.~\ref{psd_th.fig}(a)]. Consequently, the PSD of $f_o$ has a similar shape with no oscillatory peaks.

Next, we incorporate a negative auditory feedback loop into the model. We assume that the delayed vocal $f_o$ is perceived by the CNS, integrated, and then fed back to the muscle compartment. We introduce an auxiliary variable $z$ for this purpose which follows the  dynamics,
$t_z \dot{z} = -z + \mathcal{R}[F(t-\tau_a)]$, where $\mathcal{R}[\cdot]$ is a function and for simplicity we assume that vocal $f_o$ is represented by the TA force, $F(t)$.
The characteristic time, $t_z$, encompasses all processing stages of the auditory feedback system \citep{dau1996quantitative}. 
The integrated and delayed force then contributes to the excitation variable, $a(t)=a_0[1 - g_a z(t)+q\xi(t)]$, where $g_a$ is the strength of the auditory feedback. 

We consider a linear feedback first, whereby the function $\mathcal{R}$ is given by
\begin{equation}
\label{linear.eq}
\mathcal{R}(F)=\frac{F(t)-F_0}{F_0}.
\end{equation}
The model equations are
\begin{eqnarray}
\label{model3.eq}
&& \ddot{F} + \mu \dot{F} +\Omega^2 (F-F_0) + \Omega^2 g_a F_0 z=
\Omega^2 F_0 \, q \, \xi(t), \nonumber \\
&& t_z \dot{z} = -z +\frac{F(t-\tau_a)-F_0}{F_0}.
\end{eqnarray}
We analyze this 3-d order stochastic delay differential equation (DDE) first in the deterministic case, $q=0$.
To determine stability of temporal perturbations of the system about its equilibrium $F_0$, we follow the standard procedure \citep{erneux2009applied,yanchuk2017spatio}: assume exponential solution, $e^{\lambda t}$, and enter it into the DDE system (\ref{model3.eq}). This gives the following transcendental characteristic equation for the eigenvalues, $\lambda$:
\begin{equation}
\label{char1.eq}
t_z \lambda^3 + (\mu t_z + 1) \lambda^2 +(\mu+t_z\Omega^2)\lambda +\Omega^2(1+g_a e^{-\lambda \tau_a})=0.
\end{equation}
Because of the exponential term, the characteristic equation possesses infinitely many complex roots.
The equilibrium point is stable if real parts of all eigenvalues are negative. Furthermore, the existence of pairs of complex conjugate eigenvalues,
$\lambda_n=\alpha_n \pm i \beta_n$, indicates oscillatory modes with frequencies $\nu_n=\beta_n/(2\pi)$. Of importance are only a few eigenvalues with the small absolute values of their real parts, as the rest with large negative $\alpha_n$ correspond to fast decaying solutions. Roots of transcendental Eq.(\ref{char1.eq}) were found numerically and Fig.~\ref{eigen.fig}(a) exemplifies the spectrum of eigenvalues of the model (\ref{model3.eq}) for two values of the auditory feedback strength.
For $g_a=1$ real parts of all eigenvalues are negative and the equilibrium is stable, while for larger $g_a=3$ two eigenvalues possess positive real part and the system is unstable. 

Transition to instability occurs via Andronov-Hopf (AH) bifurcation when the real part of the first pair of eigenvalues crosses 0, i.e. $\text{Re} \lambda =0$ and $\text{Im} \lambda \ne 0$. Substitution of $\lambda = i \beta$ to the characteristic equation (\ref{char1.eq}) yields parametric formulas for the auditory feedback parameters $(\tau_a, g_a)$ at which the AH bifurcation occurs,
\begin{eqnarray}
&&\tau_a=\frac{1}{\beta}\,\text{Arg}(X+iY), \label{tau_hopf.eq} \\
&&X=(\mu t_z+1)\beta^2-\Omega^2, \quad Y=(\mu  + t_z \Omega^2)\beta - t_z \beta^3.\nonumber \\
&&g_a = \frac{1}{\Omega^2} \sqrt{X^2+Y^2}. \label{g_hopf.eq}
\end{eqnarray}
Figure \ref{eigen.fig}(b) shows the stability lines along which the bifurcation conditions above are satisfied. Regions below the corresponding line refer to the stability of the system. An important observation is that longer delays require less feedback strength to make the system unstable. Furthermore, the bifurcation line, $g_a(\tau_a)$, flattens out as the delay time increases, suggesting a weak dependence of the lowest frequency, $\nu_1$, on the feedback strength. This is further illustrated in Fig.~\ref{eigen.fig}(c), which compares the dependence of  $\nu_1$ on the delay time for different values of the feedback strength.
For $g_a>1$ and $\tau_a>0.2$~s the lowest frequency depends weakly on the feedback strength and is close to its value at the Andronov-Hopf bifurcation (shown by the dashed line). Finally, we note the dependence on the characteristic integration time constant, $t_z$. Fig.~\ref{eigen.fig}(b) indicates that a larger value of the feedback strength is required to bring the system to instability for a larger value of $t_z$. As expected, larger values of $t_z$ result in smaller values of frequency, $\nu_1$.

In the absence of background noise, the steady state of the linear model is just the equilibrium. When noise is taken into account
oscillatory modes become visible in the power spectrum. The PSD of the response in the stability region can be calculated as in \cite{scholl2005controlling}, i.e. by calculating the Fourier transform of the force and then ensemble averaging its square magnitude, 
$P_F (f)=\mean{|\tilde{F}(f)|^2}$. This yields,
\begin{widetext}
	\begin{equation}
	\label{psd_th.eq}
	P_F(f)=\frac{(\Omega^2 F_0 q)^2}{2 f_c} \,
	\left|\frac{1+i \omega t_z}{\Omega^2-\omega^2(1+\mu t_z) +
		i \omega(\mu+t_z(\Omega^2 - \omega^2)) + \Omega^2 g_a e^{-i\omega \tau_a}
	}\right |^2,  \quad \omega = 2\pi f.
	\end{equation}
\end{widetext}
The peaks in the PSD, shown in Fig.~\ref{psd_th.fig}(a), correspond to imaginary parts of the equilibrium's eigenvalues, $\nu_n$. The dominant peak with the lowest frequency, $\nu_1$, and lowest effective dissipation, $\alpha_1$, is followed by smaller and wider peaks centered at $\nu_2$, $\nu_3$, ... . The dominant peak at $\nu_1$ corresponds to vocal wow.
The peak position depends on the delay, $\tau_a$, as shown in Fig.~\ref{psd_th.fig}(b). Both, the real and imaginary parts of the eigenvalues decrease with the increase of the delay, resulting in sharper, lower frequency peaks.

The linear model has several drawbacks. First, the oscillations are supported solely by the physiologic tremor generated by the CNS and so are not self-sustained. Thus, the oscillation magnitude is largely determined by the parameters of the ambient noise, i.e. it's SD, $q$, and cutoff frequency, $f_c$.
Second, the system explodes when the parameters of auditory feedback ($g_a$ and $\tau_a$) are outside the stability regions bounded by the Andronov-Hopf bifurcation line, see Fig.~\ref{eigen.fig}(b). In the following section, we introduce a nonlinearity in the audiotry feedback, which limits oscillation growth in the instability region, enabling self-sustained limit cycle oscillation.

\subsection{Nonlinear model with auditory delay}
We use a sigmoid function to represent feedback nonlinearity,
\begin{equation}
\label{nonlinear.eq}
\mathcal{R}(F)=\frac{1}{2}\left[1+\tanh \left(\frac{F-F_0}{b F_0}\right) \right],
\end{equation}
where the dimensionless parameter $b$ determines the steepness and thus the sensitivity of the feedback response with respect to perturbation about unperturbed equilibrium, $F_0$. In the following we fix this parameter to the value $b=0.05$. The model's equations become,
\begin{eqnarray}
\label{model4.eq}
&& \ddot{F} + \mu \dot{F} +\Omega^2 (F-F_0) + \Omega^2 g_a F_0 z=
\Omega^2 F_0 \, q \, \xi(t), \nonumber \\
&& t_z \dot{z} = -z +\frac{1}{2}\left[1+\tanh \left(\frac{F(t-\tau_a)-F_0}{b F_0}\right) \right].
\end{eqnarray}

Stochastic DDE (\ref{model4.eq}) were solved numerically using an explicit Euler-Muryama scheme with the time step of 0.1~ms.
A $2\times 10^3$~s long sequences of vocal frequency, $f_o(t)$, were used for the PSD calculation.

In the absence of noise, $q=0$, the equilibrium force, $F_\text{eq}$, is given by
\begin{equation}
\label{equiv.eq}
F_\text{eq}+\frac{F_0}{2}\left[g_a-2+\tanh \left(\frac{F_\text{eq}-F_0}{b F_0}\right)\right]=0,
\end{equation}
whose stability is determined by the characteristic equation similar to (\ref{char1.eq}). The only difference is in the last term of (\ref{char1.eq}) where $g_a$ is replaced by $g_a \gamma$, with $\gamma$ being the derivative of $\mathcal{R}(F)$, at the equilibrium force, 
\begin{equation}
\gamma=\mathcal{R}'(F_\text{eq})=\frac{1}{2b F_0}\, \text{sech}^2 \left(\frac{F_\text{eq}-F_0}{b F_0}\right). \nonumber
\end{equation}
%

\begin{figure}[h!]
	\centering 
	\includegraphics[width=0.45\textwidth]{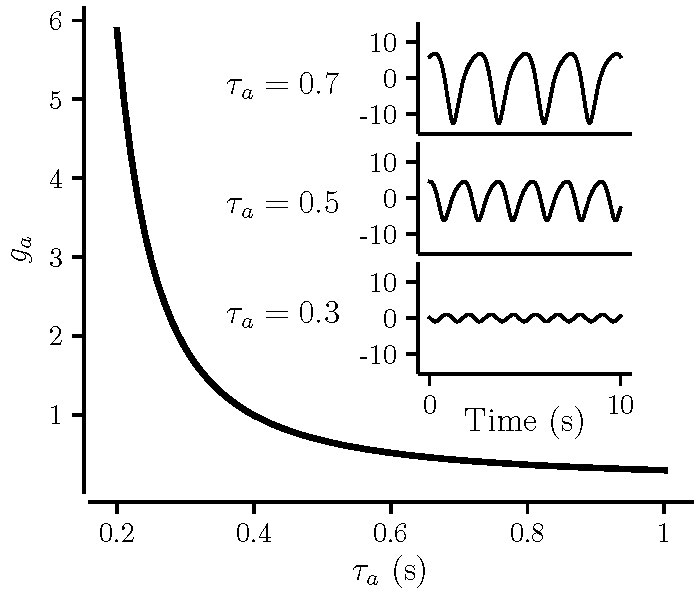}
	\caption{Andronov-Hopf bifurcation line showing the threshold value of the auditory feedback strength, $g_a$, vs the auditory delay time, $\tau_a$. Other parameters are: $t_z=0.5$~s, $b=0.05$, $F_0=1$~N, $q=0$.         
		The inset shows time traces of the vocal frequency, $f_o(t)$, for $g_a=2$ and indicated values of $\tau_a$. The vertical axes in the inset are zero-mean $f_o$ in Hz. Time traces were obtained by numerical simulations of Eqs.(\ref{model4.eq}).}
	\label{hopf.fig}
\end{figure}
Similar to the linear model, the equilibrium loses its stability via Andronov-Hopf bifurcation. However, unlike the linear model, the nonlinearity in feedback prevents unbounded growth of perturbations and instead leads to self-sustained oscillations. Figure \ref{hopf.fig}(a) shows the Andronov-Hopf bifurcation line, along which real parts of the first pair of eigenvalues vanishes, on the parameter plane $g_a$ vs $\tau_a$. The equilibrium is stable below this line, and unstable above, giving rise to a stable limit cycle.
Consequently, the vocal $f_o$, calculated according to Eq.(\ref{Fo.eq}), shows oscillations. As can be seen, the threshold feedback strength required for self-sustained oscillations decreases with the increase of the delay, $\tau_a$.
The inset in Fig.~\ref{hopf.fig} shows time traces of $f_o$ for $g_a=2$ and indicated values of the auditory delay: the amplitude and period of oscillations increase with the increase of the auditory delay. 

\begin{figure}
	\centering 
	\includegraphics[width=0.45\textwidth]{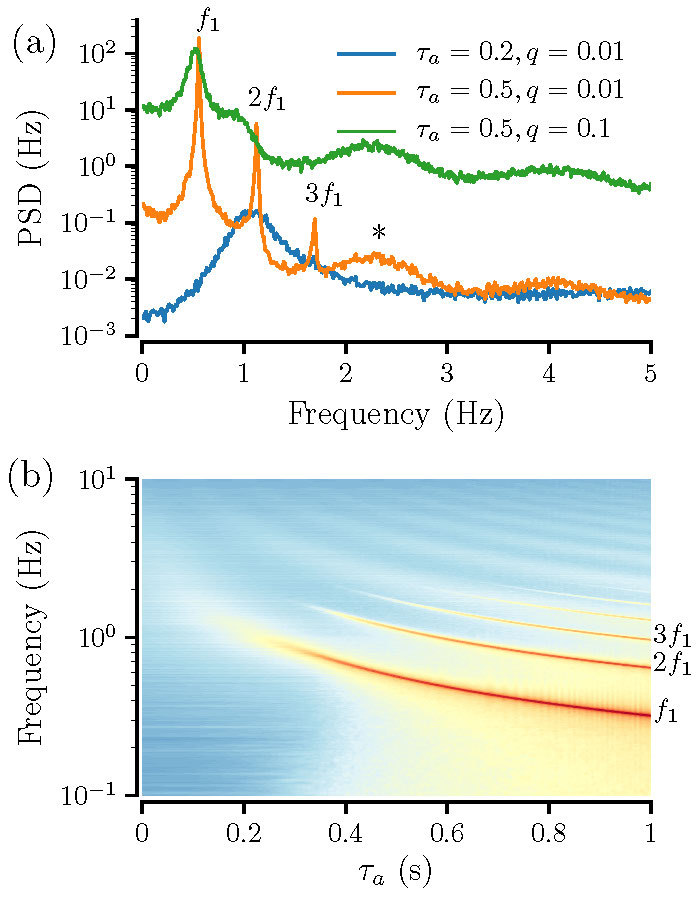}
	\caption{(color online) PSD of vocal $f_o$ for the nonlinear model (\ref{model4.eq}). 
		(a): PSD for the indicated values of the auditory delay and noise SD, $q$. The fundamental and higher harmonics are marked for $\tau_a=0.5$~s.
		The asterisk marks a broad peak at the frequency corresponding to the imaginary part of the second eigenvalue of the equilibrium, $\nu_2$.
		(b): Heat map of the PSD vs $\tau_a$. Peaks in the PSD occur at frequencies corresponding to fundamental, $f_1$, and higher harmonics, $nf_1$, marked at the right of the map.
		Other parameters: $q=0.01$, $g_a=1.5$, $t_z=0.5$~s, $F_0=1$~N.} 
	\label{psd_non.fig}
\end{figure}

With background noise taken into account, oscillation could be induced below the bifurcation line of Fig.~\ref{hopf.fig}, where the equilibrium is stable. This results in a  peak in the PSD of the vocal $f_o$, shown in Figure \ref{psd_non.fig}(a) (black line for $\tau_a=0.2$~s). In the regime of self-oscillations, i.e. above the bifurcation line of Fig.~\ref{hopf.fig}, the PSD contains a series of sharp peaks at the fundamental frequency, $f_1$, and its harmonics, $nf_1$. 
The fundamental frequency, $f_1$, corresponds to the imaginary part of the lowest eigenvalue, $\nu_1$ (cf.Fig.~\ref{psd_th.fig}). 
A broad peak corresponding to the imaginary part of the second eigenvalue, $\nu_2$, is also observed,  asterisks in Fig.~\ref{psd_non.fig}(a). 
With the increase of noise, peaks at the fundamental frequency and its higher harmonics broaden, so that for strong enough noise higher harmonics are hardly seen.
This reflects a linearization effect of noise on a nonlinear system \citep{dykman1994noise}. Consequently, the PSD structure becomes similar to that of a linear system, considered in the previous section.
For example, for $q=0.1$, shown by blue line in ~Fig.~\ref{psd_non.fig}(a), 2-nd harmonics of the fundamental can be barely seen, leaving a sharp peak at the fundamental and a much broader peak at the frequency corresponding to the second-lowest eigenvalue, $\nu_2$. 

\begin{figure}[ht]
	\centering 
	\includegraphics[width=0.45\textwidth]{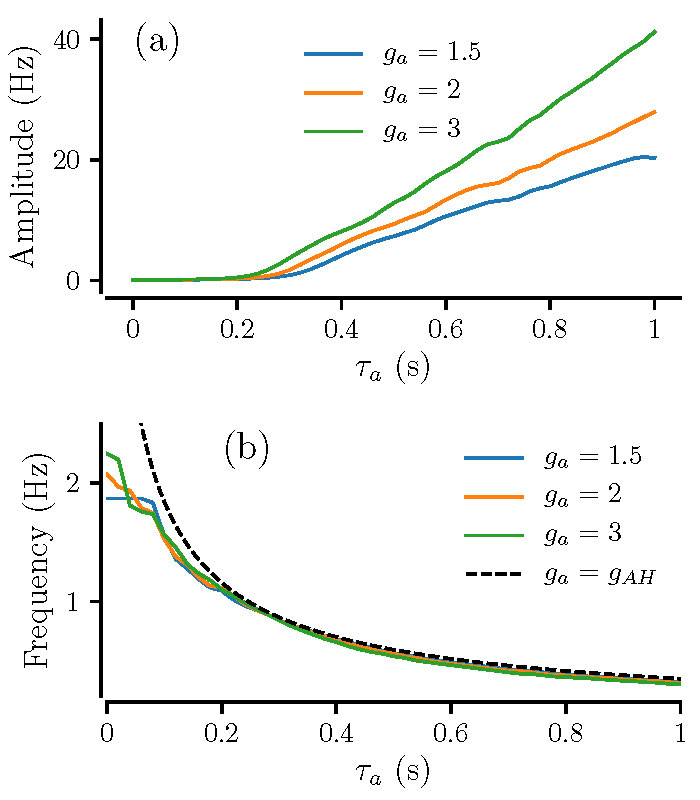}
	\caption{(color online) Amplitude and frequency of vocal $f_o$ oscillations vs delay time for the nonlinear model.
		(a) Peak-to-peak amplitude was estimated from the PSD at the dominant peak for the indicated values of the feedback strength.
		(b) Frequency of vocal $f_o$ oscillations, $f_1$, estimated by the position of the dominant peak of the PSD. 
		Other parameters: $q=0.01$, $g_a=1.5$, $t_z=0.5$~s, $F_0=1$~N.
		The dashed line shows the analytical dependence $\tau_a(f_1)$ of Eq.(\ref{tau_hopf.eq}),
		with $\beta=2\pi f_1$ and other parameters listed above.
	} 
	\label{sd_non.fig}
\end{figure}

A heat map of the PSD versus the delay time is shown in  Fig.~\ref{psd_non.fig}(b).
Similar to the linear case, the fundamental frequency decreases and its power increases with delay.
As in the noiseless case, the magnitude of vocal $f_o$ oscillations increases, while their frequency decreases with the increase of the delay time, as shown in Fig.~\ref{sd_non.fig}. 
The frequency of wow oscillations shows no significant dependence on the feedback strength. Although an exact formula for frequency, $f_1$, vs $\tau_a$ is hard to obtain, a good candidate for approximation is Eq.(\ref{tau_hopf.eq}), which essentially
gives the relation between the delay time, $\tau_a$, and the dominant frequency, $f_1=\nu_1=\beta/(2\pi)$, at the Andronov-Hopf bifurcation and approximates well the numerical results for $\tau_a>0.2$~s, as shown in Fig.\ref{sd_non.fig}(b) (dashed line). 

\subsection{Combined auditory and reflex feedback}
The original reflex resonance model demonstrates vocal $f_o$ oscillations at 5--7 Hz, corresponding to a vibrato \cite{titze2002reflex}. When delayed auditory feedback is included, we can expect quasi-periodic oscillations of vocal $f_o$ with two distinct frequencies: low wow and higher tremor. 

The full model, incorporating both the auditory and reflex feedback is described by
\begin{eqnarray}
\label{model5.eq}
&& \ddot{F}_k + \mu_k \dot{F}_k +\Omega_k^2 (F_k-F_{0k})  + \Omega^2 g_a F_{0k} z = \Omega_k^2 F_{0k} \, A_k(t),
\nonumber \\
&& t_z \dot{z} = -z +\frac{1}{2}\left[1+\tanh \left(\frac{F_2(t-\tau_a)-F_{02}}{b F_{02}}\right) \right], \nonumber \\
&& A_k(t)= q \xi(t) + g_r [\varepsilon_k(t-\tau_r)+ u_r \dot{\varepsilon}_k(t-\tau_r)].
\end{eqnarray}
In addition to these equations, the model is augmented by equations of motion of the cricoid cartilage, as discussed above. The subscript $k$ denotes CT and TA muscles, as before. Note that the negative auditory feedback (variable $z$) and noise ($\xi$ ) enters the equations of both muscles via activations, $A_k$. Parameters $F_{01}$ and $F_{02}$ are equilibrium values of the CT and TA forces, respectively, with a constant activation, i.e. when $g_a=g_r=q\equiv 0$. In the following we used $F_{01}=1.53$~N, $F_{02}=1$~N and fix the reflex delay $\tau_r=0.045$~s and the strain rate scaling constant at $u_r=0.014$ in the activation (\ref{model5.eq}).

We start with the deterministic dynamics with no background noise, $q=0$, and calculate the threshold values of the feedback loops parameters for the transition to vocal $f_o$ oscillations (periodic or quasiperiodic).
In the absence of auditory feedback, $g_a=0$, the model shows a transition to self-sustained periodic oscillations when the reflex feedback strength reaches  $g_r = g^*_r \approx5.591$, so that for $g_r > g^*_r$ the vocal $f_o$ oscillates at about 6~Hz, corresponding to the vocal vibrato.
\begin{figure}
	\centering 
	\includegraphics[width=0.45\textwidth]{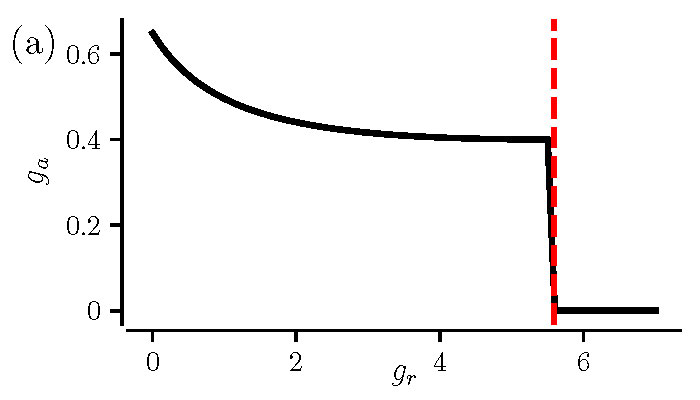}
	\vskip 0.1in
	\includegraphics[width=0.45\textwidth]{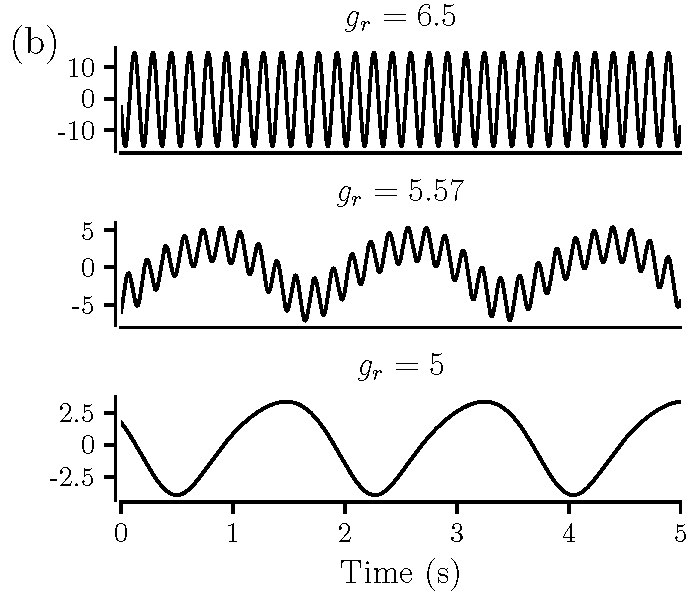}
	\caption{Effect of reflex and auditory feedback on vocal $f_o$.
		(a): Threshold value of the auditory feedback strength, $g_a$, vs the reflex feedback strength, $g_r$, for $\tau_a=0.5$~s. 
		The vertical dashed line shows the bifurcation value, $g^*_r=5.591$, at which 6~Hz vibrato oscillations emerge in the model with no auditory feedback.
		Slow wow oscillations exist above corresponding solid line; 6~Hz reflex vibrato exists to the right of  the red dashed line.
		(b): Time traces of zero-meaned $f_o(t)$ for $g_a=0.75$, $\tau_a =0.5$~s and indicated values of the reflex feedback strength, $g_r$. 
	} 
	\label{ga-gr.fig}
\end{figure}
\begin{figure*} 
	\centering
	\includegraphics[width=0.32\textwidth]{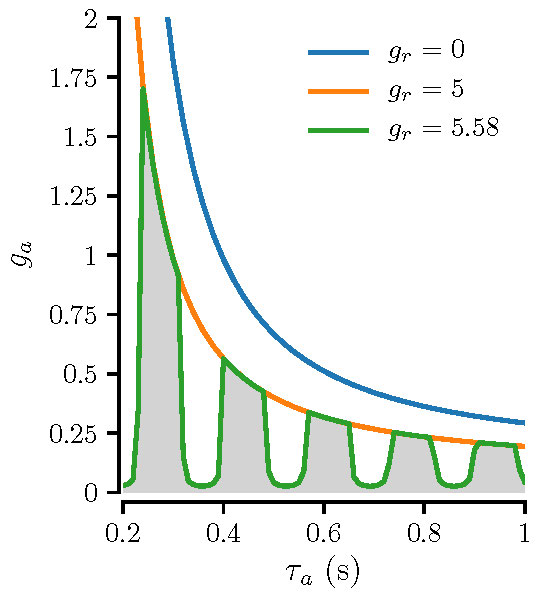}
	\includegraphics[width=0.32\textwidth]{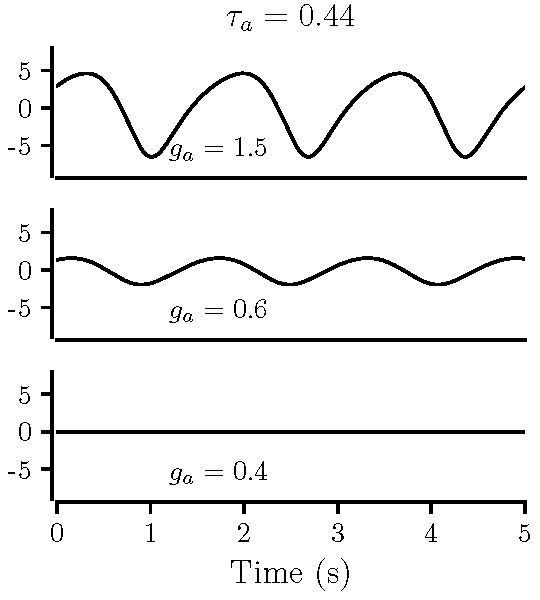}
	\includegraphics[width=0.32\textwidth]{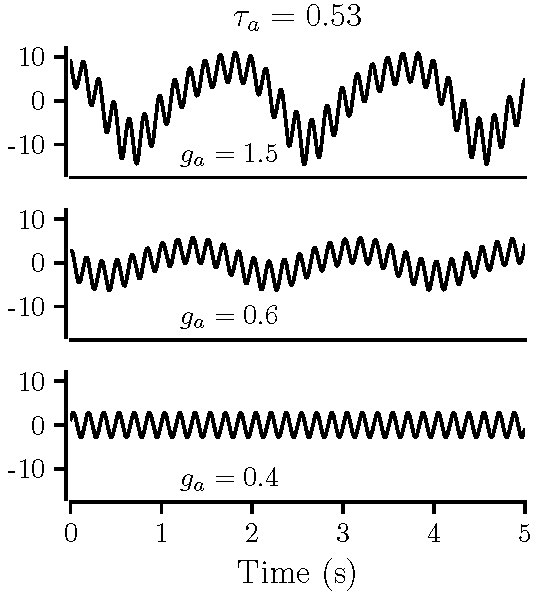}
	\caption{(color online) Effect of reflex and auditory feedback on vocal frequency $f_o$.
		Left panel: Threshold value of the auditory feedback strength, $g_a$, vs the auditory delay, $\tau_a$, for the indicated values of the reflex feedback strength, $g_r$.
		Equilibrium of the system is stable below the corresponding line, and within the shaded area for $g_r=5.57$. Self-sustained oscillations exist above relevant lines.
		Middle panel shows  transformation of $f_o$ time traces with the increase of the auditory feedback strength, when the auditory delay is set within one of the stability bands (grey shaded area) for $\tau_a=0.44$~s. Right panel shows the same, but for the auditory delay within an instability band for $\tau_a=0.53$~s. For middle and right panels the reflex feedback strength is  $g_r=5.57$.
	} 
	\label{ga-tau.fig}
\end{figure*}

With the auditory feedback on, the reflex oscillations do not show up for $g_r$ well below its bifurcation value, $g_r<g^*_r$. Nevertheless, the reflex feedback influences the onset of slow wow oscillation: the threshold value of the auditory feedback strength becomes lower as the reflex feedback strength, $g_r$ increases, as demonstrated in Fig.~\ref{ga-gr.fig}(a). In this figure, the threshold values of the feedback strength were determined numerically as the transition from the equilibrium to oscillations (periodic or quasiperiodic).
For the reflex gain close to its bifurcation value, $g^*_r$,  and the auditory feedback strength above wow oscillation threshold, $f_o(t)$ oscillations become quasiperiodic: faster reflex oscillations rides on slow wow envelope, shown in the middle panel of Fig.~\ref{ga-gr.fig}(b). With further increase of $g_r$, the reflex oscillations grow and overtake slow wow, leaving large-amplitude 6-Hz vibrato,upper panel in  Fig.~\ref{ga-gr.fig}(b).

Figure \ref{ga-tau.fig} shows the dependence of the threshold value of the auditory feedback strength, $g_a$, vs auditory delay time, $\tau_a$. For the reflex feedback strength well below its bifurcation value, $g _r< g^*_r$,  this dependence is monotonous, similar to the case of no reflex feedback: c.f. orange and blue lines in Fig.~\ref{ga-tau.fig}. For $g_a$ and $\tau_a$ above these lines, the model shows slow wow oscillations with no reflex oscillations.
Close to the bifurcation value, $g_r \succeq g^*_r$, the 
dependence $g_a(\tau_a)$ becomes non-monotonous, showing a band structure. 
In reflex oscillations, CT and TA forces are 180$^o$ phase shifted. Auditory feedback applied to CT and TA force compartments with appropriate delay diminishes the phase lag between forces, suppressing the reflex oscillations within shaded areas in Fig.  \ref{ga-tau.fig}. Outside these gaps, the model shows periodic 6-Hz reflex oscillations or quasiperiodic oscillations (for large enough auditory feedback strength), as demonstrated in the middle and right panels of Fig.~\ref{ga-tau.fig}.
\begin{figure} 
	\includegraphics[width=0.45\textwidth]{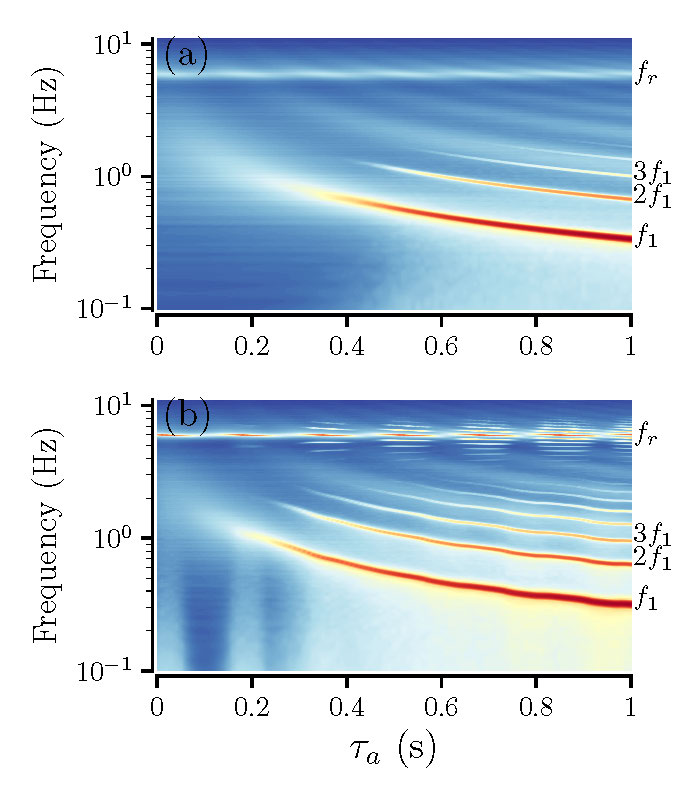}
	\caption{(color online) Heat maps of the power spectral density of vocal $f_o$ vs auditory delay time.
		(a): Value of the reflex feedback strength is set to $g_r=5$, i.e. below its bifurcation value of $g^*_r=5.591$; $g_a=0.5$.
		(b): Value of the reflex feedback gain is $g_r=5.57$, i.e. close to its bifurcation value; $g_a=1.5$.
		Noise SD is $q=0.01$.
	} 
	\label{psd_ga-tau.fig}
\end{figure}

With background stochastic activation on, the reflex and wow frequency components can be activated below threshold values of corresponding feedback strengths. Figure \ref{psd_ga-tau.fig} shows the PSD of vocal $f_o$ versus auditory delay for two sets  of feedback strengths.  For $g_r<g^*_r$  a peak at $f_r\approx 6$~Hz of reflex oscillations emerges due to random activations and co-exists with low-frequency wow, $f_1$, and its higher harmonics, $nf_1$, shown in Fig.~\ref{psd_ga-tau.fig}(a).
So that, for weak reflex feedback, the PSD of vocal $f_o$ is essentially the same as in the absence of reflex feedback, except that there is a small and broad peak at the reflex frequency, $f_r=6$Hz.
For the reflex feedback strength close or larger than $g^*_r$ and stronger auditory feedback, when the fast reflex oscillations ride on slow wow envelope, the peak at the reflex frequency possesses sidebands, $f_r \pm n f_1$, seen in Fig.~\ref{psd_ga-tau.fig} (b).

\begin{figure}
	\includegraphics[width=0.45\textwidth]{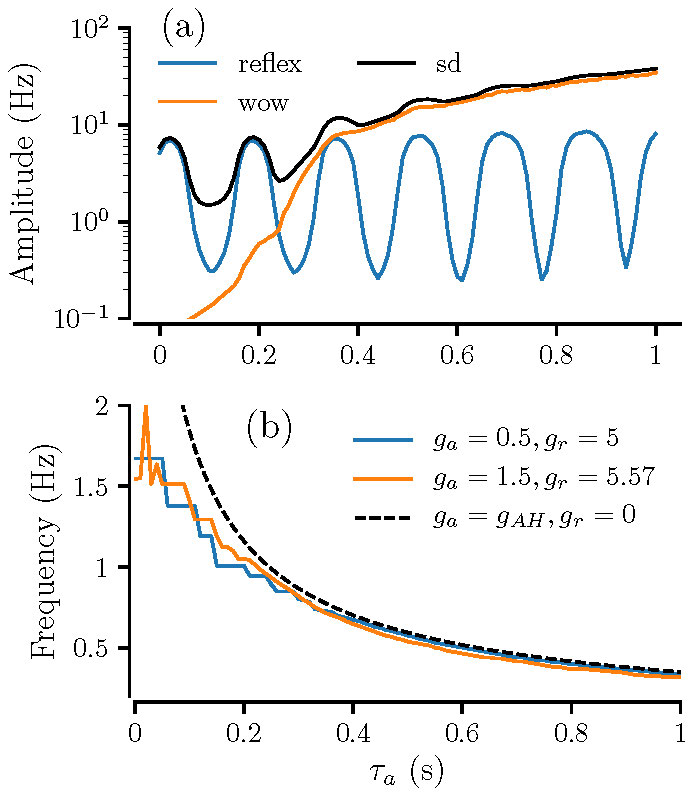}
	\caption{(color online) Amplitude and frequency of $f_o$ oscillations vs auditory delay time for the nonlinear reflex model.
		(a): Peak-to-peak amplitude of the reflex and wow oscillations of vocal $f_o$ versus auditory delay time. The overall standard deviation of $f_o$ (sd) scaled with $2\sqrt{2}$, is also shown.  The parameters are the same as in panel (b)  of Fig.~\ref{psd_ga-tau.fig}.
		(b): Frequency of wow oscillations, $f_1$, estimated by the position of the dominant peak of the PSD within 0 -- 1~Hz for the indicated values of feedback strengths.
		The dashed line shows the analytical dependence $\tau_a(f_1)$ of  Eq.(\ref{tau_hopf.eq}),  with $\beta=2\pi f_1$, also shown in Fig.~\ref{sd_non.fig}(b).
	} 
	\label{amp_gr-tau.fig}
\end{figure}

For strong reflex feedback, the amplitude of the reflex oscillations displays a non-monotonous dependence on the auditory delay time. This can be seen in the PSD heat maps as a variation of color intensity. We illustrate this further in Fig.~\ref{amp_gr-tau.fig}(a), by calculating amplitudes of the reflex and wow components from the PSD and comparing them with the overall SD of vocal $f_o$.
On this graph, minimal values of the reflex amplitude correspond to grey-shaded gaps in the bifurcation digram of Fig.~\ref{ga-tau.fig}(left panel), where the auditory feedback suppresses the reflex oscillations. For strong feedback strength non-monotonous dependence is also observed for the overall SD. Fig.~\ref{amp_gr-tau.fig}(a) shows that for large enough delay times, $\tau_a>0.3$~s, the $f_o$ time variations is domintated by wow oscillation, while the reflex is major contributor for small delays, $\tau_a<0.2$~s. 

As in the case of pure auditory feedback, the frequency of wow, $f_1$, decreases with the increase of the auditory delay time, as shown in Fig.~\ref{amp_gr-tau.fig}(b). Importantly, the dependence $f_1(\tau_a)$ follows well the analytical result Eq.(\ref{tau_hopf.eq}), when low-frequency wow dominates $f_o$ variations, i.e. for $\tau_a>0.3$~s.

\section{Proof of concept}
In line with observations by \citet{titze2002reflex}, modulation depth is dependent on feedback gain, whereas the frequency of the oscillation is primarily dependent on delay time. As these characteristics were not manifest in previous experimental findings \citep{brajot2018delay}, we conducted single-subject experiments to verify model predictions.

\subsection{Instrumentation}
Vocalizations were recorded using a head-mounted microphone (d:fine, DPA, Alleroed, Denmark), amplified and digitized with an audio interface (Fireface, RME, Haimhausen, Germany). Feedback was delayed with a digital voice processor (VoiceOne, TC Helicon, Victoria, Canada) controlled by the stimulus presentation computer using MIDI commands. Analog output from the voice processor was amplified (1202-VLZ PRO, Mackie, Seattle, WA) and presented to the subject binaurally via insert earphones (ER2, Etymotic, Elk Grove Village, IL). Both the clean and delayed signals were saved onto computer at 44100 Hz sampling rate and 16 bit quantization.

Speech intensity at the level of microphone was recorded using a digital sound level meter (SLM1, A weighting). Output was sampled at 2 Hz sampling rate and saved to file. Prior to fitting the subject with the insert earphones, the signal intensity output at the earphones was measured with a separate sound level meter (SLM2, A weighting) using a 2 c.c. coupler.

\subsection{Procedure} 
Each participant was comfortably seated in a quiet, sound-treated room. Fitted with the head-mounted microphone (5 cm distance from the mouth), the participant sustained the vowel /a/ and intensity was recorded from the microphone (SLM 1) and the external earphone (SLM 2). The gain of the microphone signal captured at the level of the audio interface (TotalMix, RME, Haimhausen, Germany) was adjusted to obtain the correspondence between interface gain and dB SPL. Based on the first few vocalizations, the participants were asked to maintain a target intensity across all trials, using visual feedback from SLM1. They had the opportunity to practice this during this first block of trials.

Participants were subsequently fitted with the insert earphones. To mask bone conduction, a speech-weighted noise was played to subjects binaurally. 
To verify that the speech-weighted noise effectively masked bone-conducted speech, participants reported whether they could hear their voice with the microphone off and masking noise present. The participants then completed an initial block of vowel prolongations as feedback gain was increased incrementally at a fixed delay of 300 ms. Subsequent blocks were carried out with varying delay and fixed feedback gain. Feedback gain and delay settings are described for each participant.

\subsection{Data Analysis}
Fundamental frequency contours were extracted from each audio recording (Matlab Audio Toolbox, MathWorks, Natuck, NJ) and visually truncated so as not to include large peaks in $f_o$ at the beginning of trials due to hard phonation onsets. Any linear trend in each signal was removed before calculating the waveform standard deviation and peak-to-peak amplitude. The power spectrum density was then calculated and all peaks at least 16\% of the maximum were automatically identified. The largest peak between 0.3 and 3.5 Hz was selected as the principal wow frequency.


\subsection{Results}
Participant 1 was a 23 year-old male, non-musician. He passed a hearing screen with thresholds below 20 dB HL for frequencies between 250 and 8000 Hz in both ears. With insert earphones in place and no masking noise present, his maximum phonational frequency range extended from 108 Hz to 245 Hz. The masking noise presented on all subsequent experimental trials was set to 73 dB SPL.
\begin{figure}
	\centering 
	\includegraphics[width=0.45\textwidth]{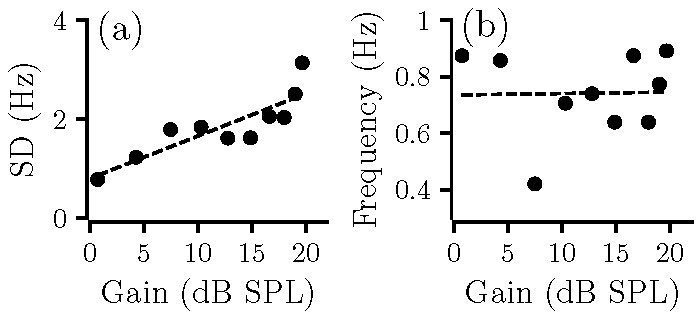}
	\caption{Vocal $f_o$ standard deviation (a) and peak frequency (b) vs feedback gain with delay $\tau = 0.3$~s for Participant 1. The dashed lines represent linear least-squares fits.}
	\label{P1_gain.fig}
\end{figure}
On the first block of trials, the audio signal fed back to the participant was delayed by 300 ms and its gain increased in increments of 3 dB. The duration of analyzed signals ranged from 6.8 to 8.4 s ($M=7.6, SD=0.6$). With visual feedback, the participant managed to maintain vocal intensity between 71.4 and 74.6 dB SPL ($M=73.2, SD=1.0$). With increasing gain, the standard deviation of each $f_o$ contour increased by 0.9 Hz/dB 
[Fig. \ref{P1_gain.fig}(a)]. The corresponding increase in peak-to-peak amplitude was 0.5 Hz/dB. 
The primary frequency of oscillations ranged from 0.42 to 0.89 Hz ($M=0.74, SD=0.15$), with no appreciable change with increasing feedback gain [Fig.~\ref{P1_gain.fig}(b)].
\begin{figure}
	\centering 
	\includegraphics[width=0.45\textwidth]{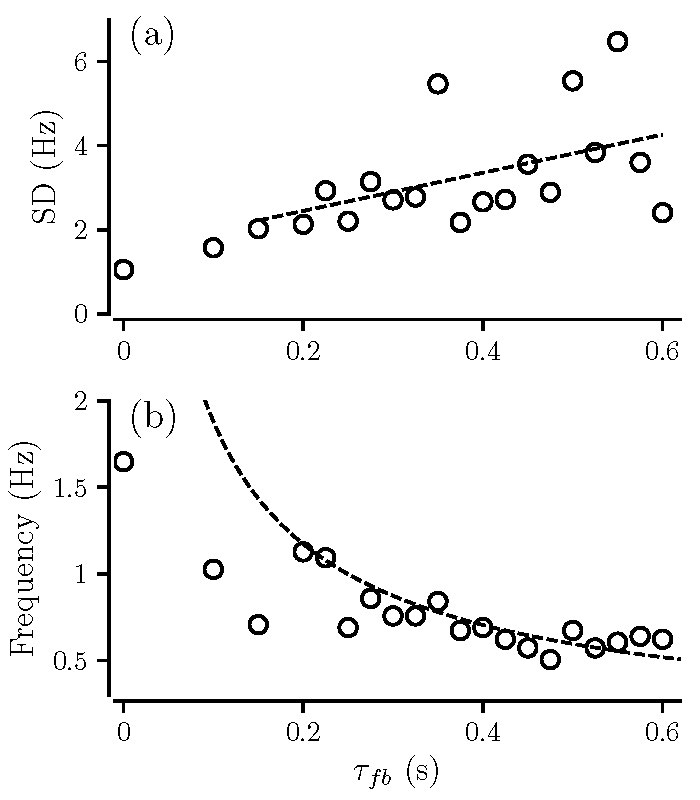}
	\caption{Vocal $f_o$ standard deviation (a) and peak frequency (b) vs feedback delay for Participant 1, with a feedback gain of +18 dB SPL.  On panel (a) the dashed line represents the linear fit; on panel (b) the dashed line shows least-squares fit with  Eq.(\ref{tau_hopf.eq}).}
	\label{P1_delay.fig}
\end{figure}

Participant 1 completed a second block of trials with a feedback gain of +18 dB and delays ranging from 0 to 600 ms. The duration of analyzed signals ranged from 6.3 to 8.9 s ($M = 7.9, SD = 0.7$). Peak vocal intensity ranged from 76 to 80.2 dB SPL ($M = 78.2, SD = 0.9$), with a 1 to 2 dB mean increase above 450 ms delay. Mean vocal $f_o$ increased almost 30 Hz when auditory feedback was delayed, from 142 Hz at 0 ms delay to a mean 171 Hz ($SD = 5$, range: 158-178 Hz) for delays $\geq$ 100 ms.

The primary frequency, $f_1$, of $f_o$ oscillations ranged from 0.24 to 1.9 Hz ($M=0.85, SD=0.31$) and decayed with increasing delay time [Fig.~\ref{P1_delay.fig}(b)]. 
We fit peak modulation frequency, $f_1$, against delay $\tau$, setting $\beta=2\pi f_1$ with $\mu$, $\Omega^2$,  $t_z$ as fitting parameters in (\ref{tau_hopf.eq}) and applying a nonlinear least-square fit function (lsqcurvefit; Optimization Toolbox, Matlab 2019a).
Biomechanical time constants $t_s$ and $t_\sigma$ were then calculated from the fitting parameters $\mu$ and $\Omega^2$. 
The resulting fit approximated the experimental data moderately well for delays above 150 ms, with an  $R^2=0.886$ and $SSE=0.157$ [Fig.~\ref{P1_delay.fig}(b)].
The parameters recovered from the fit were $t_z=0.507$s, $t_\sigma=0.008$s, $t_s=0.042$s,  comparable to the values reported by \cite{titze2002reflex} and included in the present model.
\begin{figure}
	\centering 
	\includegraphics[width=0.45\textwidth]{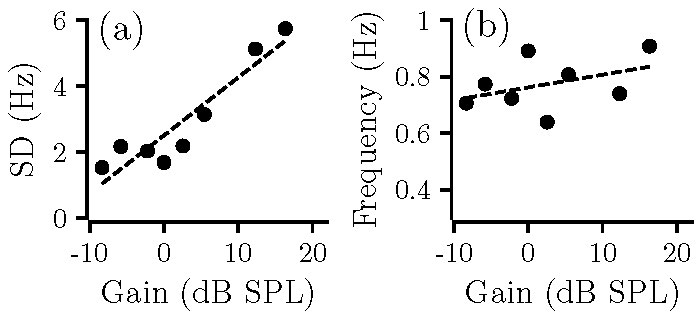}
	\caption{Vocal $f_o$ standard deviation (a), and peak frequency (b), vs feedback gain with delay $\tau = 0.3$~s for Participant 2. The dashed lines represent linear least-squares fits.}
	\label{P2_gain.fig}
\end{figure}

Participant 2 was a 65 year-old female diagnosed with relapsing-remitting type Multiple Sclerosis 18 years prior.
With insert earphones in place and no masking noise present, her maximum phonation frequency range extended from 154 to 400 Hz,. The masking noise presented on all subsequent experimental trials was set to 73 dB SPL.
She completed a first block of trials with a fixed delay of 300 ms and an incremental increase in feedback gain from -9 to 16 dB SPL. The duration of the analyzed $f_o$ contours ranged from 6.7 to 10 s ($M = 8.7, SD = 1$). Mean peak intensity was 75 dB SPL ($SD = 2$) and mean vocal $f_o$ 247 Hz ($SD = 1.62$). Fig.~\ref{P2_gain.fig}(a)  shows $f_o$ standard deviation as a function of feedback gain, rising at a rate of 0.18 Hz/dB (equivalent peak-to-peak amplitude: 1 Hz/dB). The predominant frequency of oscillation shown in Fig.~\ref{P2_gain.fig}(b) remained relatively stable across trials ($M = 0.77, SD = 0.09$, range = 0.64 -- 0.91 Hz).

\begin{figure}
	\centering 
	\includegraphics[width=0.45\textwidth]{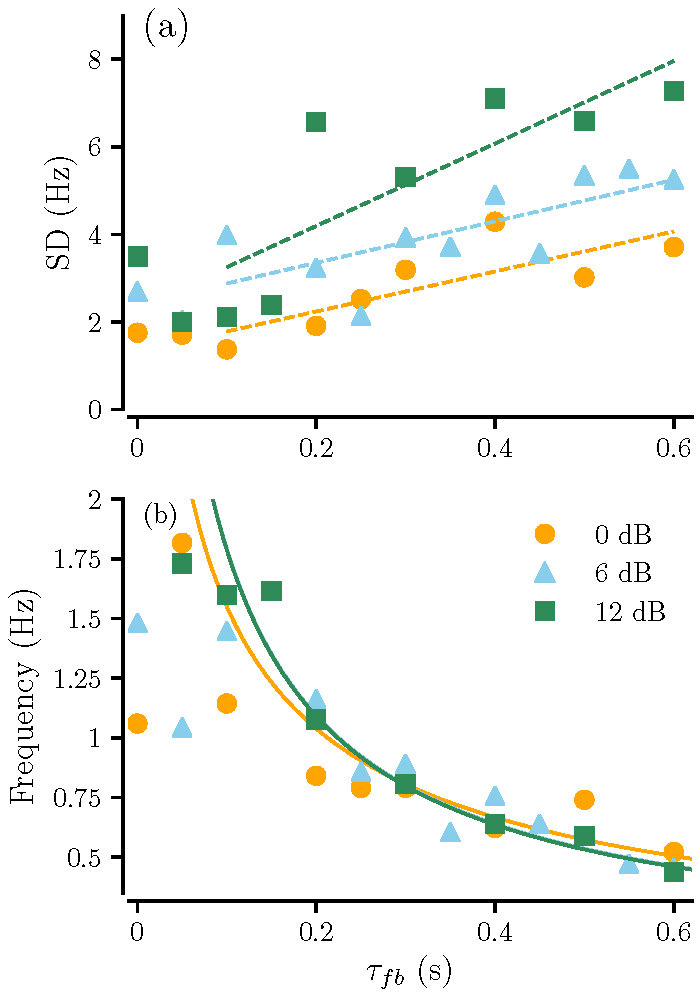}
	\caption{(color online) Vocal $f_o$ standard deviation (a) and the wow frequency (b) vs feedback delay at three indicated values of feedback gains for Participant 2. 
		On panel (a) dashed lines  show linear fit for corresponding sets of feedback gains; 	on panel (b) solid lines show least square fit with  Eq.(\ref{tau_hopf.eq}).
	}
	\label{P2_delay.fig}
\end{figure}

Participant 2 then repeated three blocks of trials with set feedback gains of 0, 6, and 12 dB SPL above the intensity recorded at the microphone. Across all conditions, the analyzed signal durations ranged from 6.5 to 9.8 s ($M = 8.6, SD = 0.9$). Mean peak intensity was 75.3 dB SPL ($SD = 2.6$) and mean vocal $f_o$ 242 Hz ($SD = 6.84$).
The standard deviation of $f_o$ contours increased with delay at approximately 4, 5 and 9 Hz/s with gains of 0, 6 and 12 dB respectively [Fig.~\ref{P2_delay.fig}(a)]. The corresponding peak-to-peak amplitude increase was 26, 19 and 54 Hz/s, which is qualitatively similar to the modeling results shown in Fig.~\ref{sd_non.fig}(a).

\begin{figure*}
	\centering 
	\includegraphics[width=0.45\textwidth]{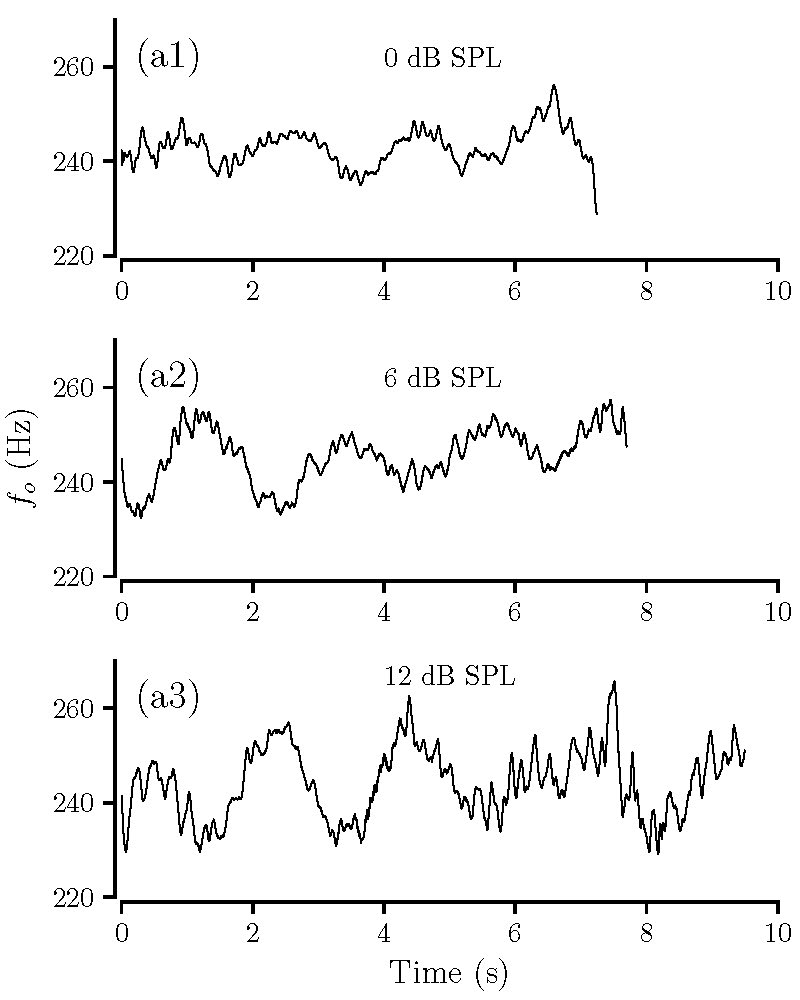}
	\includegraphics[width=0.45\textwidth]{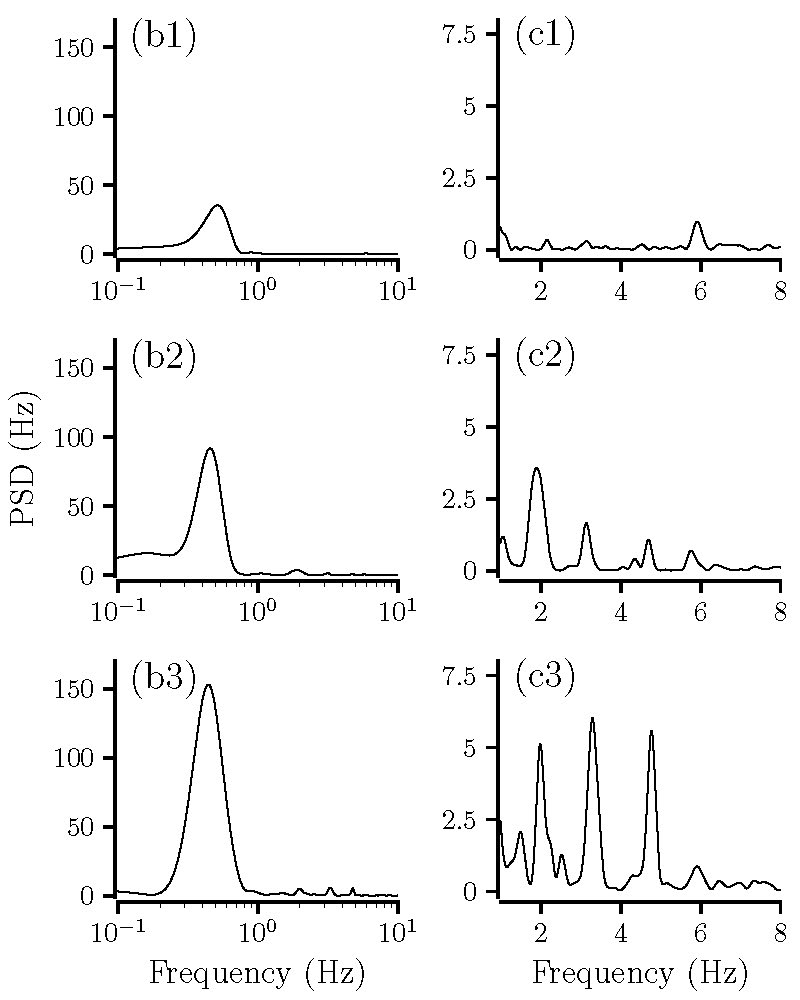}
	\caption{Time series of vocal $f_o$ (a1--a3) for the indicated values of the auditory feedback gain at $\tau=0.6$~s. (b1--b3): PSDs corresponding to panels (a). (c1--c3): The same PSDs on the expanded 1--8~Hz frequency range.
	}
	\label{P2_tremor.fig}
\end{figure*}

The predominant frequency $f_1$ identified from the $f_o$ power spectra decayed with increasing delay across all gain conditions [Fig.~\ref{P2_delay.fig}(b)]. 
Similar to  Participant 1, decays were well fitted by Eq.(\ref{tau_hopf.eq}) for delays $ \ge 200$~ms.
Furthermore, fitted curves were indistinguishable for gains of 6 and 12 dB.
The fit across all gain conditions reached an $R^2=0.961$, with an $SSE=0.072$.
The recovered parameters were $t_z=0.944$s, $t_\sigma=0.008$s, $t_s=0.042$s. 
Tremor was intermittently perceptible in this participant's vocalizations. Unfortunately, the number of productions elicited were deliberately limited in order to mitigate vocal fatigue. Too few data were available to identify a clear interaction between wow and tremor. We were also unable to directly control for changes in reflex parameters.
We did however observe an overall increase in 2-8 Hz spectral energy for delays $\ge 200$~ms as feedback gain increased. This was characterized by an increase in the number of peaks in that range, rather than an increase in the original 6 Hz tremor observed in the 0 dB feedback gain condition.
Example $f_o$ contours and respective PSDs for 600 ms delay are displayed in Fig. \ref{P2_tremor.fig}.

\section{Discussion}
Adding an auditory feedback loop to the reflex resonance model proposed by \citet{titze2002reflex}, we have shown that a delay can effectively induce a slow modulation of fundamental frequency. 
In terms of system dynamics, the addition of non-linear negative feedback results in the emergence of an otherwise absent low-frequency oscillatory mode. 
The frequency of this vocal wow drops  as delay is increased, following the dependence described by Eq.(\ref{tau_hopf.eq}).
As is the case for vibrato in the original model, wow extent (modulation depth) is primarily dependent on feedback gain whereas the frequency of oscillation depends on delay. 
Model output with both reflex and auditory feedback loops in place shows interesting interactions between wow and tremor, moreover. Tremor predominates when reflex gain is very high. When reflex gain is near bifurcation, different patterns emerge depending on auditory feedback parameters. With high auditory gain, tremor is superimposed on vocal wow. With delays in auditory feedback, tremor extent waxes and wanes depending on the delay.
Furthermore, reflex feedback can heighten the system's sensitivity to auditory feedback, even when little or no tremor is observed in the power spectrum.

The correspondence between model approximations and experimental data is encouraging, despite noise inherent in the experimental approach and the difficulty identifying true peaks in such short signals. Importantly, the model allows us to recover the delay and time constants based on peak modulation frequency in vocal $f_o$ (following Eq.\ref{tau_hopf.eq}). This has clear clinical implications, as it may be possible to quantify the impact of a disease on the control mechanisms important for voice and speech.

There are, of course, a number of caveats to consider.
First, although we note a strong dependence of wow frequency on delay magnitude on the one hand, and of wow extent on feedback gain on the other, these are not fully independent. There does appear to be a weak dependence of wow frequency on auditory feedback gain, in particular for delays below 200 ms.
This translates to increased variability for human subject data (cf. Fig 11b, 14b), which may be further complicated by the interaction between feedback loops as reflex gain increases.

Secondly, we introduced an auditory feedback loop with dynamics and non-linearity that, while consistent with current models of auditory processing, was heuristically motivated; that is, we deliberately sought the simplest model that would allow us to recover specific parameters.
Given the complexity of pitch processing in the auditory system \citep{dau1996quantitative},
the model may benefit from the incorporation of additional auditory processing parameters. Additional control loops, such as auditory efferents, may also prove insightful, if not helpful, albeit at the expense of increased complexity.

Third, we should note that we do not distinguish between gain and delay parameters within the auditory system and those being experimentally manipulated. The distinction is unimportant for the current analysis, but this may become relevant when considering central nervous system diseases that result in delayed conduction times before or after posited non-linearities in the auditory system.

Lastly, with respect to the full model, we provided a rather brief account for the nonlinear interaction of wow and tremor oscillatory modes. In particular we conducted direct simulations with a fixed value for the reflex delay, without detailed bifurcation analysis of transition to quasiperiodicity \cite[which can be done, e.g.  using a parameter continuation technique;][]{engelborghs2002numerical}. An interesting and important question is how the variations of both the auditory and reflex time delays affect $f_o$ oscillatory patterns. 

We surrender these lines of inquiry to future research.

\begin{acknowledgements}
	AN thanks Serhiy Yanchuk and Micahel Zaks for valuable discussions and  acknowledges support by the Lobachevsky
	University of Nizhny Novgorod through the Russian Science Foundation grant 14-41-000440.
\end{acknowledgements}

\end{document}